\input amstex
\documentstyle{amsppt}
%
%
%
\nopagenumbers
\nologo
\def\negskp{\hskip -2pt}
\def\compos{\,\raise 1pt\hbox{$\sssize\circ$} \,}
\def\divr{\operatorname{div}}
\def\rot{\operatorname{rot}}

\accentedsymbol\ty{\tilde y} \pagewidth{450pt} \pageheight{650pt}
\leftheadtext{S.~F.~Lyuksyutov, R.~A.~Sharipov, G.~Sigalov,
P.~B.~Paramonov} \rightheadtext{Exact analytical solution for
electrostatic field\dots} \topmatter
\title
Exact analytical solution for electrostatic field produced by
biased atomic force microscope tip dwelling above
dielectric-conductor bi-layer
\endtitle
\author
S.~F.~Lyuksyutov, R.~A.~Sharipov, G.~Sigalov, P.~B.~Paramonov
\endauthor
\address Departments of Physics, Chemistry and Polymer Engineering, Ayer Hall, 213,
The University of Akron, Akron, OH 44325
\endaddress
\email sfl\@physics.uakron.edu
\endemail
\address Mathematics Department, Bashkir State University,
Frunze street 32, 450074 Ufa, Russia
\endaddress
\email \vtop to 20pt{\hsize=280pt\noindent
R\_\hskip 1pt Sharipov\@ic.bashedu.ru\newline
r-sharipov\@mail.ru\vss}
\endemail
\urladdr
http:/\negskp/www.geocities.com/r-sharipov
\endurladdr
\abstract
   An exact analytical solution based on the method of images has been
obtained for the description of the electrostatic field in the
system consisting of atomic force microscope (AFM) tip, dielectric
and conductor. The solution provides a step towards quantitative
modelling of the AFM-assisted electrostatic nanolithography in
polymers.
\endabstract
\endtopmatter
\loadbold
\TagsOnRight
\document
\head 1. Introduction
\endhead
   Atomic Force Microscopy is an important tool for nanoscale modifications
in metals, semiconductors, and soft condensed matter. Polymers
suggest clear advantage with respect to the other materials in
such fields as data storage and sacrificial patterning. Recently,
an electrostatic nanolithography based on AFM \cite{1,2} suggested
a way of patterning nanostructures in thin (10-50 nm) dielectric
films coated onto metal substrate. Although the physical
description of the process based on electrostatic attraction of
softened polymer toward the tip in strong non-uniform electric
field has apparently been understood, the mathematical model has
not been developed until this time.

   The goal of this Letter is to derive the exact analytical
solution for the spatial distribution of the electric field and
potential in the system consisting of a biased AFM tip, dielectric
polymer film, and a conductor using the method of images. This
solution is required for phenomenological description of the
tip-polymer interaction and electrostatic pressure formed inside
the softened dielectric material. The model we introduce differs
from that considered in \cite{3} due to the presence of a
conductive substrate. \hskip -2em

\head
2. Model description.\\
Field equations and boundary conditions
\endhead

A conceptual presentation of the system comprising conductive AFM
tip and bi-layer consisting of the polymer film and conductive
substrate is shown in Figure~1. An electrically biased AFM tip is
presented as an equipotential spherical surface, modelled by a
charge $Q$ in its center. An opposite charge of planar density
$\sigma$ is distributed over the surface of the dielectric
film.\par

\includegraphics{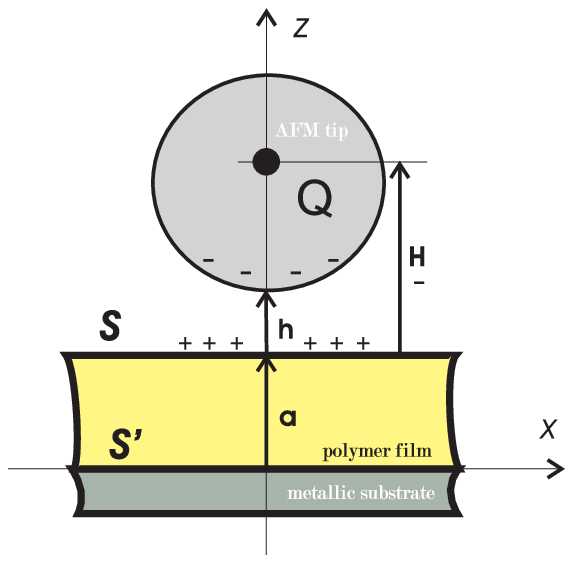} \vskip 167pt\hbox{\kern
55pt{\it Figure~1 }\hss} \vskip -174pt

\vskip 50pt

   The electric field $\bold E=\bold E(\bold r)$, produced by $Q$
and $\sigma$, should satisfy the following electrostatic equations
\cite{4}:
$$
\xalignat 2 &\hskip -2em \rot\bold E=0, &&\divr\bold E=0.\quad
\tag2.1
\endxalignat
$$
The electric field inside the bulk conductor equals zero: $\bold
E(\bold r)=0$ for $z<0$. The field is perpendicular to the
conductor - dielectric interface $S'$:
$$
\hskip -2em \bold E_{\sssize\parallel}\,\hbox{\vrule height 8pt
depth 8pt width 0.5pt}_{\,\bold r\in S'}=0. \tag2.2
$$

The tangential component of $\bold E$ is continuous at the
interface $S$, while normal components of the electric field
$\bold E$ and the electric displacement vector $\bold D$ undergo a
discontinuity related to $\sigma$:
$$
\xalignat 2 &\hskip -2em \bold E_{\sssize\parallel}\,\hbox{\vrule
height 8pt depth 8pt width 0.5pt}_{\,\bold r\in S-0}= \bold
E_{\sssize\parallel}\,\hbox{\vrule height 8pt depth 8pt width
0.5pt}_{\,\bold r\in S+0}, &&\bold D_{\sssize\perp}\,\hbox{\vrule
height 8pt depth 8pt width 0.5pt}_{\,\bold r\in S-0}+\sigma= \bold
D_{\sssize\perp}\,\hbox{\vrule height 8pt depth 8pt width
0.5pt}_{\,\bold r\in S+0}. \tag2.3
\endxalignat
$$
For the linear dielectric media, $\bold D$ is linearly
proportional to $\bold E$:
$$
\hskip -2em
\bold D=\cases\epsilon_0\,\varepsilon_{\sssize\text{air}}
\,\bold E&\text{\ \ for \ }z>a\text{\ \ above $S$},\\
\epsilon_0\,\varepsilon_{\sssize\text{pol}}\,\bold E &\text{\ \
for \ }0<z<a\text{\ \ below $S$}.
\endcases
\tag2.4
$$
Here $\varepsilon_{\sssize\text{air}}$ and
$\varepsilon_{\sssize\text{pol}}$ are the dielectric constants of
air and polymer film respectively. The second boundary condition
\thetag{2.3} can be rewritten as
$$
\hskip -2em \varepsilon_{\sssize\text{pol}}\,\bold
E_{\sssize\perp}\,\hbox{\vrule height 8pt depth 8pt width
0.5pt}_{\,\bold r\in S-0} +\frac{\sigma}{\epsilon_0}=
\varepsilon_{\sssize\text{air}}\,\bold
E_{\sssize\perp}\,\hbox{\vrule height 8pt depth 8pt width
0.5pt}_{\,\bold r\in S+0}. \tag2.5
$$
The combination of \thetag{2.5} and the first part of
\thetag{2.3},
$$
\xalignat 2 &\hskip -2em \bold E_{\sssize\parallel}\,\hbox{\vrule
height 8pt depth 8pt width 0.5pt}_{\,\bold r\in S-0}= \bold
E_{\sssize\parallel}\,\hbox{\vrule height 8pt depth 8pt width
0.5pt}_{\,\bold r\in S+0},
&&\varepsilon_{\sssize\text{pol}}\,\bold
E_{\sssize\perp}\,\hbox{\vrule height 8pt depth 8pt width
0.5pt}_{\,\bold r\in S-0} +\frac{\sigma}{\epsilon_0}=
\varepsilon_{\sssize\text{air}}\,\bold
E_{\sssize\perp}\,\hbox{\vrule height 8pt depth 8pt width
0.5pt}_{\,\bold r\in S+0}, \tag2.6
\endxalignat
$$
along with \thetag{2.2}, provide the boundary conditions for the
differential equations \thetag{2.1}.

\hskip -2em

\head 3. Basic (auxiliary) electrostatic field in air
\endhead

   In the general case, the charge density distribution $\sigma$
on the polymer surface depends on the tip's motion with respect to
the polymer surface. The axially symmetric $\sigma$ is given by
$$
\hskip -2em \sigma(r)=\frac{-Q'\,k\,H}{2\,\pi\,
(k^2\,H^2+r^2)^{3/2}},\text{ \ where \ }r=\sqrt{x^2+y^2}. \tag3.1
$$
The total charge, distributed on the polymer surface, equals
$-Q'$:
$$
2\pi\int\limits^\infty_0\sigma(r)\,r\,dr=-Q'.
$$
The degree of spatial concentration of the surface charge density
can be varied adjusting the parameter $k>0$ in \thetag{3.1}.\par

   It is convenient to use the following auxiliary construction in
the derivation of the electrostatic field distribution for the
system of our interest. If we consider the system of the point
charge $Q$ and charge density $\sigma(r)$ at the absence of the
dielectric and conductor, the electrostatic field is given by
$$
\pagebreak
\hskip -2em
\bold E_{\sssize\text{bas}}(\bold r)=\cases
\dsize\frac{Q}{4\,\pi\,\epsilon_0\,\varepsilon_{\sssize\text{air}}}
\,\frac{\bold r-\bold r_{\sssize Q}}{|\bold r-\bold r_{\sssize Q}|^3}
-\frac{Q'}{4\,\pi\,\epsilon_0\,\varepsilon_{\sssize\text{air}}}
\,\frac{\bold r-\bold r'_{\sssize Q}}
{|\bold r-\bold r'_{\sssize Q}|^3}\text{ \ below \ }S,
\\
\dsize\frac{Q}{4\,\pi\,\epsilon_0\,\varepsilon_{\sssize\text{air}}}
\,\frac{\bold r-\bold r_{\sssize Q}}{|\bold r-\bold r_{\sssize Q}|^3}
-\frac{Q'}{4\,\pi\,\epsilon_0\,\varepsilon_{\sssize\text{air}}}
\,\frac{\bold r-\bold r''_{\sssize Q}}{|\bold r-\bold r''_{\sssize Q}|^3}
\text{ \ above \ }S.
\endcases
\tag3.2
$$
Here $\bold r_{\sssize Q}=\bold a+\bold H$, $\bold r_{\sssize
Q'}=\bold a+k\,\bold H$ and $\bold r_{\sssize Q''}=\bold
a-k\,\bold H$ are radii-vectors of the actual charge $Q$, and the
virtual (image) charges $-Q'$, placed below the interface $S$.
Below, we modify the basic electrostatic field \thetag{3.2} to
describe the case when the dielectric - conductor bi-layer is
placed below $S$.

\hskip -2em

\head 4. Solution of the original electrostatic problem
\endhead

   The problem is to construct the actual electric field
distribution $\bold E(\bold r)$ for the case when an external
field $\bold E_{\sssize\text{bas}}(\bold r)$ applies to the
dielectric - conductor bi-layer. To solve this problem we use the
method of images \cite{4}. A discrete set of vectors is introduced
in the following form:
$$
\hskip -2em \bold a_n=2\,n\,\bold a,\qquad n=0,1,2,\ldots,\infty.
\tag4.1
$$
Reflection operator acts
$$
\hskip -2em
\bold r=\Vmatrix x\\ y\\ z\endVmatrix\longrightarrow
\overline{\bold r}=\Vmatrix x\\ y\\ -z\endVmatrix.
\tag4.2
$$
The reflection operation is denoted by the bar above the vector.
We seek solution for the electric field inside the polymer film in
the form
$$
\bold E(\bold r)=\sum^{\infty}_{n=0}\alpha_n\,
\bold E_{\sssize\text{bas}}(\bold r-\bold a_n)
-\sum^{\infty}_{n=0}\alpha_n\,
\overline{\bold E}_{\sssize\text{bas}}(\overline{\bold r}-\bold a_n).
\tag4.3
$$
The form of the sum \thetag{4.3} satisfies the boundary condition
\thetag{2.2} for the conductor - dielectric interface $S'$. For a
point at $S'$, $\overline{\bold r}=\bold r$, and $\bold
E_{\sssize\text{bas}}(\bold r-\bold a_n)= \bold
E_{\sssize\text{bas}}(\overline{\bold r}-\bold a_n)$. Hence,
$$
\hskip -2em \bold E^{\sssize\parallel}_{\sssize\text{bas}}(\bold
r-\bold a_n)\, \hbox{\vrule height 8pt depth 8pt width
0.5pt}_{\,\bold r\in S'} =\overline{\bold E}\vphantom{\bold
E}^{\sssize\parallel}_{\sssize \text{bas}}(\overline{\bold
r}-\bold a_n)\, \hbox{\vrule height 8pt depth 8pt width
0.5pt}_{\,\bold r\in S'}, \tag4.4
$$
and continuity of the tangential component of the electric field
is satisfied.\par
   Now, the components of the electric field
$\bold E_{\sssize\parallel}$ and $\bold E_{\sssize\perp}$ at the
interface $S$ of the polymer film can be found. For the point
$\bold r\in S$, $\overline{\bold r}=\bold r -2\,\bold a$; thus,
$\overline{\bold r}-\bold a_n=\bold r-\bold a_{n+1}$ and the
following relationship holds:
$$
\bold E_{\sssize\text{bas}}(\overline{\bold r}-\bold a_n)\,
\hbox{\vrule height 8pt depth 8pt width 0.5pt}_{\,\bold r\in S}
=\bold E_{\sssize\text{bas}}(\bold r-\bold a_{n+1})\,
\hbox{\vrule height 8pt depth 8pt width 0.5pt}_{\,\bold r\in S}.
\tag4.5
$$
The following notations are introduced for convenience:
$$
\xalignat 2
&\hskip -2em
G^{\sssize\parallel}_n=\bold E^{\sssize\parallel}_{\sssize
\text{bas}}(\bold r-\bold a_n)\,
\hbox{\vrule height 8pt depth 8pt width 0.5pt}_{\,\bold r\in S},
&&G^{\sssize\perp}_n=\bold E^{\sssize\perp}_{\sssize
\text{bas}}(\bold r-\bold a_n)\,
\hbox{\vrule height 8pt depth 8pt width 0.5pt}_{\,\bold r\in S}.
\tag4.6
\endxalignat
$$
Using \thetag{4.3}, \thetag{4.5}, \thetag{4.6} and \thetag{4.2},
the following two expressions for normal and tangential field
components can be written:
$$
\align
&\hskip -2em
\bold E_{\sssize\parallel}\,\hbox{\vrule height 8pt depth 8pt
width 0.5pt}_{\,\bold r\in S-0}=\alpha_0\,G^{\sssize\parallel}_0
+\sum^{\infty}_{n=0}(\alpha_{n+1}-\alpha_n)\,G^{\sssize\parallel}_{n+1},
\tag4.7\\
&\hskip -2em \bold E_{\sssize\perp}\,\hbox{\vrule height 8pt depth
8pt width 0.5pt}_{\,\bold r\in S-0}=\alpha_0\,G^{\sssize\perp}_0
+\sum^{\infty}_{n=0}(\alpha_{n+1}+\alpha_n)\,G^{\sssize\perp}_{n+1}.
\tag4.8
\endalign
$$
The electric field above the boundary $S$ is presented as
$$
\hskip -2em
\bold E(\bold r)=\bold E_{\sssize\text{bas}}(\bold r-\bold a_0)
+\beta_{-1}\,\overline{\bold E}_{\sssize\text{bas}}(\overline{\bold r}
-\bold a_{-1})+\sum^{\infty}_{n=0}\beta_n\,\overline{\bold E}_{\sssize
\text{bas}}(\overline{\bold r}-\bold a_n).
\tag4.9
$$
Combining \thetag{4.5}, \thetag{4.6} and \thetag{4.2} with
\thetag{4.9} we find
$$
\align
&\hskip -2em
\bold E_{\sssize\parallel}\,\hbox{\vrule height 8pt depth 8pt
width 0.5pt}_{\,\bold r\in S+0}=\left(1
+\beta_{-1}\right)G^{\sssize\parallel}_0+
\sum^{\infty}_{n=0}\beta_n\,G^{\sssize\parallel}_{n+1},
\tag4.10\\
&\hskip -2em
\bold E_{\sssize\perp}\,\hbox{\vrule height 8pt depth 8pt
width 0.5pt}_{\,\bold r\in S+0}=G^{\sssize\perp}_0-\beta_{-1}\,
G^{\sssize\perp}_0-\sum^{\infty}_{n=0}\beta_n\,G^{\sssize
\perp}_{n+1},
\tag4.11
\endalign
$$
Considering \thetag{4.7}, \thetag{4.8} and \thetag{4.10},
\thetag{4.11} simultaneously, together with the boundary
conditions \thetag{2.6}, the relationships for series coefficients
can be found:
$$
\hskip -2em
\left\{
\aligned
&1+\beta_{-1}=\alpha_0,\\
&\varepsilon_{\sssize\text{air}}\,(1-\beta_{-1})
=\varepsilon_{\sssize\text{pol}}\,\alpha_0.
\endaligned
\right.
\tag4.12
$$
and
$$
\hskip -2em
\left\{
\aligned
\beta_n&=\alpha_{n+1}-\alpha_n,\\
-\varepsilon_{\sssize\text{air}}\,\beta_n&=
\varepsilon_{\sssize\text{pol}}\,(\alpha_{n+1}+\alpha_n),
\endaligned
\qquad n=0,1,\ldots,\infty. \right. \tag4.13
$$
Note, that the basic field \thetag{3.2} undergoes discontinuiety
at the interface $S$. Two different values of $G^{\sssize\perp}_0$
are obtained from \thetag{4.8} and \thetag{4.11}:
$$
\xalignat 2 &G^{\sssize\perp}_0(+)=\bold E^{\sssize\perp}_{\sssize
\text{bas}}(\bold r-\bold a_0)\, \hbox{\vrule height 8pt depth 8pt
width 0.5pt}_{\,\bold r\in S+0}, &&G^{\sssize\perp}_0(-)=\bold
E^{\sssize\perp}_{\sssize \text{bas}}(\bold r-\bold a_0)\,
\hbox{\vrule height 8pt depth 8pt width 0.5pt}_{\,\bold r\in S-0}.
\endxalignat
$$
The difference of these two values is related to the surface
charge density $\sigma$:
$$
\hskip -2em G^{\sssize\perp}_0(+)-G^{\sssize\perp}_0(-)=
\frac{\sigma}{\epsilon_0\, \varepsilon_{\sssize\text{air}}}.
\tag4.14
$$
   Solving \thetag{4.12} and \thetag{4.13} for the series coefficients
$\alpha_n$ and $\beta_n$, we can find the recurrent relationships
$$
\xalignat 2 &\hskip -2em
\alpha_{n+1}=-\frac{\varepsilon_{\sssize\text{pol}}-
\varepsilon_{\sssize\text{air}}}
{\varepsilon_{\sssize\text{pol}}+\varepsilon_{\sssize\text{air}}}
\,\alpha_n, &&\beta_n=-\frac{2\,\varepsilon_{\sssize\text{pol}}}
{\varepsilon_{\sssize\text{pol}}+\varepsilon_{\sssize\text{air}}}\,
\alpha_n, \tag4.15
\endxalignat
$$
and
$$
\xalignat 2
&\hskip -2em
\alpha_0=\frac{2\,\varepsilon_{\sssize\text{air}}}
{\varepsilon_{\sssize\text{pol}}+\varepsilon_{\sssize\text{air}}},
&&\beta_{-1}=-\frac{\varepsilon_{\sssize\text{pol}}-
\varepsilon_{\sssize\text{air}}}{\varepsilon_{\sssize\text{pol}}
+\varepsilon_{\sssize\text{air}}}.
\tag4.16
\endxalignat
$$
From \thetag{4.15} and \thetag{4.16}, we finally obtain the
explicit expressions for the series coefficients:
$$
\hskip -2em
\aligned
&\alpha_n=\left(-\frac{\varepsilon_{\sssize\text{pol}}
-\varepsilon_{\sssize\text{air}}}
{\varepsilon_{\sssize\text{pol}}+\varepsilon_{\sssize\text{air}}}
\right)^n\,
\frac{2\,\varepsilon_{\sssize\text{air}}}
{\varepsilon_{\sssize\text{pol}}+\varepsilon_{\sssize\text{air}}},\\
\vspace{2ex}
&\beta_n=-\left(-\frac{\varepsilon_{\sssize\text{pol}}-
\varepsilon_{\sssize\text{air}}}{\varepsilon_{\sssize\text{pol}}
+\varepsilon_{\sssize\text{air}}}\right)^n\,
\frac{4\,\varepsilon_{\sssize\text{pol}}\,\varepsilon_{\sssize\text{air}}}
{(\varepsilon_{\sssize\text{pol}}+\varepsilon_{\sssize\text{air}})^2}
\endaligned
\tag4.17
$$
Introducing
$$
\eta=\frac{\varepsilon_{\sssize\text{pol}}-
\varepsilon_{\sssize\text{air}}} {\varepsilon_{\sssize\text{pol}}+
\varepsilon_{\sssize\text{air}}}, \tag4.18
$$
and noting that
$$
\xalignat 2
\frac{\varepsilon_{\sssize\text{air}}}
{\varepsilon_{\sssize\text{pol}} +\varepsilon_{\sssize\text{air}}}
=\frac{1-\eta}{2}, &&\frac{\varepsilon_{\sssize\text{pol}}}
{\varepsilon_{\sssize\text{pol}} +\varepsilon_{\sssize\text{air}}}
=\frac{1+\eta}{2}, \tag4.19
\endxalignat
$$
we have
$$
\xalignat 2 &\hskip -2em
 \alpha_0=1-\eta, &&\beta_{-1}=-\eta,
\tag4.20\\
&\hskip -2em
\alpha_n=(-\eta)^n\,(1-\eta),
&&\beta_n=-(-\eta)^n\,(1-\eta^2).
\tag4.21
\endxalignat
$$
Finally, the electrostatic field inside the polymer, and above the
polymer film is given by
$$
\bold E(\bold r)=(1-\eta) \sum^{\infty}_{n=0}(-\eta)^n\!\left(
\bold E_{\sssize\text{bas}}(\bold r-\bold a_n) -\overline{\bold
E}_{\sssize\text{bas}} (\overline{\bold r}-\bold a_n)\right),
\tag4.22
$$
$$
\hskip -2em
\gathered
\bold E(\bold r)=\bold E_{\sssize\text{bas}}(\bold r-\bold a_0)
-\eta\,\overline{\bold E}_{\sssize\text{bas}}(\overline{\bold r}
-\bold a_{-1})-\\
-(1-\eta^2)\sum^{\infty}_{n=0}(-\eta)^n\, \overline{\bold
E}_{\sssize \text{bas}}(\overline{\bold r}-\bold a_n).
\endgathered
\tag4.23
$$

\hskip -2em

\head 5. Induced charges
\endhead
   The electric field polarizes polymer film with the degree of
polarization described through the induced density of the dipole
moment $\bold P$. Vector $\bold P$ is given by
$$
\hskip -2em
\aligned
&\bold P_{\sssize\text{air}}=\epsilon_0\,
(\varepsilon_{\sssize\text{air}}-1)\,
\bold E_{\sssize\text{air}},\\
&\bold P_{\sssize\text{pol}}=\epsilon_0\,
(\varepsilon_{\sssize\text{pol}}-1)\,
\bold E_{\sssize\text{pol}}.
\endaligned
\tag5.1
$$
The bulk and the surface charge distributions appear as a result
of polarization induced by the electric field:
$$
\xalignat 2 &\hskip -2em \rho=-\divr\bold P, &&\sigma=\bold
P\cdot\bold n, \quad \tag5.2
\endxalignat
$$
for the volume and surface charge densities. Here $\bold n$ is the
unit vector normal to the interface $S$. The constancy of the
dielectric constants $\varepsilon_{\sssize\text{pol}}$ and
$\varepsilon_{\sssize\text{air}}$ yields $\divr\bold D=0$ and
$\divr\bold E=0$. This implies no volume electric charges inside
the polymer film. Using Gauss' theorem, the electric field due to
the surface charge on the interface $S$ is presented as
$$
\hskip -2em \bold E_{\sigma}\,\hbox{\vrule height 8pt depth 8pt
width 0.5pt}_{\,\bold r\in S+0}- \bold E_{\sigma}\,\hbox{\vrule
height 8pt depth 8pt width 0.5pt}_{\,\bold r\in S-0}=
\frac{\sigma}{\epsilon_0}\,\bold n. \tag5.3
$$
This field must be subtracted from the expression for the net
electrostatic field, given by \thetag{4.22} and \thetag{4.23}, in
order to exclude self-action when calculating tension forces and
pressure associated with the field. The resulting function $\bold
E(\bold r)-\bold E_{\sigma}(\bold r)$ is continuous on $S$, and
its value on $S$ is
$$
\hskip -2em (\bold E-\bold E_{\sigma})\,\hbox{\vrule height 8pt
depth 8pt width 0.5pt}_{\,\bold r\in S}= \frac{\bold E(\bold
r)}{2}\,\hbox{\vrule height 12pt depth 8pt width 0.5pt}_{\,\bold
r\in S+0}+ \frac{\bold E(\bold r)}{2}\,\hbox{\vrule height 12pt
depth 8pt width 0.5pt}_{\,\bold r\in S-0}=\bold
E_{\sssize\text{av}}. \tag5.4
$$
The traction acting on the upper surface of the polymer film,
related to the stress induced by the electrostatic field, is
$$
\hskip -2em \bold T=\sigma \bold E_{\sssize\text{av}}. \tag5.5
$$
The surface density of polarization charges is expressed through
the electric field inside the polymer film, \thetag{4.22},
according to \thetag{5.2} and \thetag{5.3}, as
$$
\hskip -2em \sigma_{\sssize\text{pol}}=\epsilon_0\,
(\varepsilon_{\sssize\text{pol}}-1)\, \bold
E^{\sssize\perp}_{\sssize\text{pol}}. \tag5.6
$$

\hskip -2em

\head
Summary
\endhead
   The exact analytical solution for a spatial distribution of the
electrostatic field in the system consisting of the electrically
biased AFM tip and the dielectric - conductor bi-layer has been
derived. The solution has been developed in the series form using
the method of images. The expressions have been obtained for the
charge density and traction of the electrostatic stress at the
polymer surface. Represented results provide a step towards the
quantitative understanding of the AFM-assisted electrostatic
nanolithography in polymers.

\hskip -2em


\Refs \widestnumber\no{2} \ref\no 1\by Lyuksyutov~S.~F.,
Vaia~R.~A., Paramonov~P.~B., Juhl~S., Waterhouse~L., Ralich~R.~M.,
Sigalov~G., Sancaktar~E.\paper Electrostatic nanolithography in
polymers using atomic force microscopy\jour Nature Materials 2
(2003) 468-472
\endref
\ref\no 2\by Lyuksyutov~S.~F., Paramonov~P.~B., Juhl~S.,
Vaia~R.~A. \paper Amplitude-modulated electrostatic
nanolithography in polymers based on atomic force microscopy\jour
Appl. Phys. Lett. 83 (2003) 4405-4407
\endref
\ref\no 3\by Engel~A., Friedrichs~R.\paper On the electromagnetic
force on a polarizable body\jour cond-mat/0105265 in Electronic
Archive {\bf http:/\negskp/arXiv.org}
\endref
\ref\no 4\by Landau~L,~D., Lifshitz~E.~M.\book Electrodynamics of
continuous media\publ Pergamon Press, NY\yr 1960
\endref
\endRefs
\enddocument
\end